# Spatial Resolution of Local Field Potential Signals in Macaque V4

Armin Najarpour Foroushani[1], Sujaya Neupane[2], Pablo De Heredia Pastor[1], Christopher C. Pack[2**], and Mohamad Sawan[1,3,4**]

[1]Polystim Neurotech Lab., Electrical Engineering Dept., Polytechnique Montreal, Quebec, Canada
[2]Montreal Neurological Institute and Hospital, McGill University, Montreal, Canada
[3]School of Engineering, Westlake University, Hangzhou 310024, China
[4]Institute of Advanced Study, Westlake Institute for Advanced Study, Hangzhou 310024, China

Email: armin.najarpour-foroushani@ polymtl.ca.
** Both authors contributed equally to this work

## Abstract

A main challenge for the development of cortical visual prostheses is to spatially localize individual spots of light, called phosphenes, by assigning appropriate stimulating parameters to implanted electrodes. Imitating the natural responses to phosphene-like stimuli at different positions can help in designing a systematic procedure to determine these parameters. The key characteristic of such a system is the ability to discriminate between responses to different positions in the visual field. While most previous prosthetic devices have targeted the primary visual cortex, the extrastriate cortex has the advantage of covering a large part of the visual field with a smaller amount of cortical tissue, providing the possibility of a more compact implant. Here, we studied how well ensembles of Multiunit activity (MUA) and Local Field Potentials (LFPs) responses from extrastriate cortical visual area V4 of a behaving macaque monkey can discriminate between two-dimensional spatial positions. We found that despite the large receptive field sizes in V4, the combined responses from multiple sites, whether MUA or LFP, has the capability for fine and coarse discrimination of positions. We identified a selection procedure that could significantly increase the discrimination performance while reducing the required number of electrodes. Analysis of noise correlation in MUA and LFP responses showed that noise correlations in LFP responses carry more information about the spatial positions. Overall, these findings suggest that spatial positions could be localized with patterned stimulation in extrastriate area V4.

## 1. Introduction

A main challenge for the development of cortical visual prostheses is to generate spatially localized spots of light, called phosphenes, to restore perception of visual scenes [1-3]. In practice, the characteristics of these phosphenes depend on the location and stimulation parameters of the implanted electrodes in the visual cortex [4-9]. Although much progress has been made in generating and modulating individual phosphenes, no experimental study has been reported to restore a visual scene by systematic control over the locations of multiple phosphenes. Previous studies however, showed that the percepts induced by electrical stimulation are similar to the percepts elicited by visual stimuli matching the receptive field locations of neurons at the microstimulated site [8, 10-12]. This suggests that reconstruction of the natural responses of visual cortex to phosphene-like stimuli at different positions in the visual field can assist at designing a systematic procedure to control the position of multiple phosphenes by tuning the stimulation parameters. The key characteristic of such a system then, would be its capability to discriminate between responses to phosphene-like visual stimuli at different spatial positions.

Multielectrode recordings are necessary to obtain both the spatial and temporal sampling properties required for the representation of position [13, 14]. However, the majority of previous multielectrode recordings and stimulation studies in the visual cortex of monkeys have been performed in V1 [6, 8, 15-23], which is quite large relative to standard recording arrays. Therefore, a multielectrode array in V1 can often sample only a tiny region of visual space. In contrast, extrastriate visual areas generally contain retinotopic maps that are physically smaller, while the receptive fields are much larger than those in the primary visual cortex (V1) [24-26]. This provides the opportunity to sample a larger region of visual space, albeit with reduced spatial resolution, using standard devices such as Utah arrays [27, 28]. In particular, extrastriate area V4 offers an opportunity to recover the location of static visual stimuli, as it contains a retinotopic map of visual space [28, 29] and responds well to stimuli of low to moderate complexity [30, 31]. However, it remains to be seen how well the neuronal ensemble activity provided by V4 can support discrimination of different positions in the visual space.

While chronically implanted recording arrays could provide the necessary information for stimulus localization, it is well known that the ability to resolve individual neurons with these devices declines over time [32]. However, local field potentials (LFPs) are more durable and can often be measured reliably for years after the implantation of multi-electrode arrays [33, 34]. They are low frequency fluctuations [35, 36] generated by the synaptic current flow in neural ensembles [37-39]. They reflect subthreshold activity at larger spatial and temporal scales than single unit recording [40, 41], and they can be modulated by microstimulation [42-45]. Importantly, previous work has shown that the LFPs in V4 provide retinotopic information comparable to that obtainable with neuronal activity [28].

Since the receptive fields in V4 are large and have extensive spatial overlap, each position in visual space projects onto a large population of neurons. The accuracy of these receptive fields in discriminating fixed positions in the visual space is remained to be addressed. Chen et al. showed that overlapping receptive fields provide sufficient spatial precision in area MT [27]. However, these results were obtained using the multiunit spiking activity (MUA). Here, we chronically implanted Utah electrode arrays with 96 electrodes to study the spatial discrimination in area V4. We recorded both LFP and MUA activity from area V4 while individual probe stimuli were presented to the macaque monkey. The probes were squares of 2° visual angle with similar characteristics to the phosphenes [4, 46-49]. We determined the spatial precision with which these signals could discriminate fixed positions in the visual field. We proposed and compared two procedures to reduce the number of electrodes used for the discriminations. Then, we characterized the coding strategy made by the decoder in discrimination of probes at different positions. We also examined whether band passed LFPs can provide a better spatial resolution than the broad band LFP. In addition, we identified the structure of noise correlation as a source of information about spatial position and showed their strengths as a function of cortical distances.

## 2. Materials and Methods

*2.1 Electrophysiological recordings and signal pre-processing*

All aspects of the experiment were approved by the Animal Care Committee of the Montreal Neurological Institute and were conducted in agreement with regulations of Canadian Council of Animal Care. The recording procedure has been described before [28, 50]. Briefly, a sterile surgical procedure was performed under general anesthesia to implant a headpost and a 10×10 Utah Microelectrode Array (Utah array; Blackrock Microsystems) with 1 mm electrodes with 400 µm spacing in area V4 of macaque monkeys (*Macaca fascicularis*). Area V4 was identified based on its anatomical landmarks and stereotactic coordinates [51]. After recovering from the surgery, the monkey was seated in a primate chair (Crist Instruments) and trained and rewarded to make visually guided saccades.

The data were collected by simultaneously recording wideband extracellular signal using a standard data acquisition system (Plexon Multichannel Acquisition Processor System) with a sampling rate of 10 kHz over all 96 channels of Utah Microelectrode Array. Preliminary signal-processing and custom setting of preamplifier were performed as explained previously [50]. The raw signals were band-pass filtered between 500 and 4000 Hz to form the MUA signals. LFP signals were obtained by removing action potential signals from the raw signal using a Bayesian spike removal algorithm [52], band-pass filtering of the resulted signal between 0.2-150 Hz and then down-sampling it to 500 Hz [53].

*2.2 Visual stimulation*

The experimental procedure has been described previously [53]. Briefly, visual stimuli were presented at a 78 cm viewing distance on a semi-transparent screen with refresh frequency 75Hz. The screen has covered 80×50° visual angle. Visual stimuli were white square probes with 2° width and a distance of 4° center-to-center from their neighbours in both the vertical and horizontal directions. The luminance of the probes was 22.5 cd m$^{-2}$ on a dark background with less than 0.01 cd m$^{-2}$ luminance. Each probe was chosen randomly among 100 locations on a 10×10 grid placed in the lower left quadrant of the visual field. The grid location was determined to cover the retinal eccentricity of the receptive fields of all the recorded neurons; these were centered within the central 40° of the lower left visual hemifield (see Figure 1(a)).

On each trial, the monkey fixated its gaze on a small red dot of 0.5° diameter on the top right of the grid. If the monkey did not maintain gaze within 2.5° of the point, the trial was aborted. After 500 ms of fixation, a randomly chosen probe stimulus was flashed for 25 ms separated by a 500 ms blank screen. Each probe location on the grid was repeated 14-21 times in a pseudorandom order which led to the total of 1769 trials. Each trial ended with a liquid reward given to the monkey for fixating on the red dot.

*2.3 LFP and MUA analysis*

All the analyses were performed in Python 3.6. The recorded activity for each trial was analyzed starting 350 ms before and ending 350 ms after stimulus onset. Multiunit activity (MUA) at each electrode site was presented as spike counts computed in nonoverlapping time windows of width 25 ms. The evoked MUA response of each electrode on each trial was defined as the total spike count over a certain time window. We defined three time windows, all beginning 50 ms after the stimulus onset; wide (150 ms), medium (50 ms), and narrow (25 ms). We created separate response datasets based on each measure for each window. For each dataset, the responses then were z-scored across all the trials for each electrode independently.

Given the previous results [28], the LFP amplitude has better localization than the instantaneous LFP power in area V4. Besides, the negative amplitude of the broadband LFP signals (0.2-150 Hz) is helpful in discriminating stimuli at different positions [53]. Therefore, we defined the LFP response as the mean amplitude of the broadband LFP signal over time windows same as those defined for MUA. Like MUA analysis, the responses for each dataset were z-scored across all the trials for each electrode separately.

To confirm that the information about stimulus position was not restricted to certain frequency bands, we applied fourth order Butterworth FIR filter to the broadband LFPs recorded 350 ms before to 350 ms after stimulus onset. We filtered these signals over five frequency bands: theta (4-8 Hz), alpha (8-12 Hz), beta (12-30 Hz), gamma (30-50 Hz), and high gamma (50-80 Hz) independently. The delta frequency band (0.5-4 Hz) was not used in the analysis, as it does not capture responses over each of the three response windows. We then calculated the responses and created response datasets for each band passed signal, similar to the broadband LFP.

*2.4 Spatial receptive field map*

The spatial receptive field map of each recording site was obtained by averaging the z-scored responses over trials with the same probe positions. This formed a receptive field map that was then smoothed with a Gaussian filter of standard deviation 0.8 and linearly interpolated in two dimensions (10×10).

We also fitted a two-dimensional Gaussian model to the receptive field maps separately through least-squares minimization:

$$G(A, \mu_X, \mu_Y, \sigma_X, \sigma_Y, d) \sim G(X, Y) = A e^{-(\frac{(X-\mu_X)^2}{2\sigma_X^2} + \frac{(Y-\mu_Y)^2}{2\sigma_Y^2})} + d \qquad (1)$$

The Gaussian model had six parameters: maximum response $A$, centre position ($\mu_X, \mu_Y$), SD ($\sigma_X, \sigma_Y$), and bias $d$. We used X-Y positions of each probe on the grid (the fovea location as the reference) as the input to the model and the mean response to each probe as the output. The Gaussian model was then fitted and the receptive fields were presented as the ellipses of full width at half height of the fitted Gaussian with diameters $D_X = 2\sigma_X \sqrt{-2\ln(\frac{1}{2} - \frac{d}{2A})}$, $D_Y = 2\sigma_Y \sqrt{-2\ln(\frac{1}{2} - \frac{d}{2A})}$. The diameter of each receptive field was reported as the average of $D_X$ and $D_Y$.

*2.5 Selecting informative electrodes*

To interpret the discrimination performance between pairs of positions and have an unbiased sample for the subsequent analysis, we removed recording electrodes that were not informative about the probe positions on the grid. We applied univariate feature selection to all the MUA and LFP datasets by computing ANOVA F-values. Electrodes showing p<0.05 were selected for subsequent analyses. Separate analyses including all the electrodes resulted in poorer discrimination performance (not shown).

*2.6 Discrimination analysis*

Discrimination analysis was performed by classifying ensemble responses for every pairs of probe positions on the grid. Discrimination between pairs of probes when at least one is not within the receptive field of the responsive electrodes is difficult to interpret. To collect probes that elicited responses, we calculated the mean z-score across electrodes and trials at each probe position [27]. For MUA, positions for which this value was positive and for LFP, positions for which it was negative were included in the discrimination analysis.

Support vector machines (SVMs) have frequently been used in previous studies to quantify the discriminability of population activity for pairs of stimuli [27, 54-56]. SVMs are effective in high dimensional spaces and in cases where the number of dimensions is greater than the number of samples. We therefore used SVMs on the MUA and LFP responses separately to discriminate pairs of probe positions. Given that each probe position was repeated 14-21 times, the data set for every pair contained between 28 and 42 samples. Support vector machines build a classification hyperplane by maximizing the distance to the nearest data points of any class. Given training vectors $x_i \in \mathbb{R}^p, i = 1, \ldots, n$ and targets $y \in \{1, -1\}^n$, SVM solves:

$$\min_{w,b,\xi} \frac{1}{2} w^T w + \frac{C}{2} \sum_{i=1}^{n} \xi_i^2 \qquad (2)$$

Subject to $y_i(w^T \varphi(x_i) + b) \geq 1 - \xi_i, \ \xi_i \geq 0, i = 1, \ldots, n$

where, $C > 0$ is the upper bound, function $\varphi$ maps the training samples into higher dimensional space [57], and $K(x_i, x_j) = \varphi(x_i)^T \varphi(x_j)$ is the kernel function. We separately applied linear SVM and SVM with Radial Basis Function (RBF) kernel defined as $K(x_i, x_j) = e^{-\gamma \|x_i - x_j\|^2}$. In this setting, the hyperparameter $C$ trades off correct classification of training samples against maximization of the margin (complexity of the decision function). Thus, it behaves as a regularization parameter. The hyperparameter $\gamma$ of the RBF kernel determines the influence of a single training sample.

We randomly shuffled the order of the trials in the training set to make sure that the training set was the representative of the overall distribution of the data. Since the number of repetitions differed across probe positions, we adjusted the SVM weights in inverse proportion to the class frequencies in the input data to keep the balance of the learned weights. We used a 5-fold cross validation procedure to prevent overfitting: all the trials were divided into 5 groups of trials, called folds, of equal sizes (if possible). The training was performed using 4 folds, and the remaining fold was used for validation. This procedure was repeated 5 times by using each of the folds once as the left-out fold. The validation score was determined as the mean of the 5 scores. Because of the imbalance in the number of repetitions of probes, we used stratified mode of 5-fold cross validation which preserves the relative class frequencies in each train and validation fold [58]. Similar results were obtained with 10-fold stratified cross validation (not shown).

Not every pair of probes can be discriminated with the same values of the hyperparameters. Discrimination between each pair of probes required a new adjustment in the hyperparameters to prevent overfitting. To this end, we applied Grid Search [59] over $C$ and $\gamma$ to determine the combination with the highest stratified 5-fold cross-validation score (mean cross-validated score). This score was reported as the validation performance for discrimination of each probe pair.

Because of the imbalance in the number of repetitions of the probes and also because two probes (negative and positive targets) were equally important in each single discrimination, we used the area under the Receiver Operating Characteristic (ROC) curve to measure validation performance [60]. ROC curve plots true positive rate (TPR) or sensitivity versus false positive rate (FPR) or one minus specificity. Each prediction result represents one point in the ROC space. The area under this curve (AUC) is a measure of performance (in classification) that ranges in value from 0 to 1.

In discrimination of two positions, Linear SVM assigns a weight to each electrode based on its importance. The positive weights support the hypothesis that the stimulus is at a target position, while the negative weights support the hypothesis that the stimulus is at another position. We independently normalized the SVM weights corresponding to each discrimination. Since the L2 norm was used in the formulation of the SVM here, we defined the importance of each site in each discrimination as the square of its weight [61]. The overall importance of each electrode was calculated as the mean of the importance values over all the discriminations (we only considered discriminations with separation distances of <15°). We sorted the recording sites based on their importance to find the most and least important sites over all the discriminations.

*2.7 Probes separation distances, eccentricity, and cortical magnification factor*

The positions of probes on the grid do not have the same distance from the fovea center (fixation point). Discrimination of pairs of probes with the same separation distance on the grid, thus, will not give the same level of performance. Therefore, it is important to place the discrimination performance in the context of the retinal eccentricity of the receptive fields, taking into account cortical magnification factor. Cortical magnification factor is the ratio of distance between two points on the cortex (in millimeters) divided by distance between their corresponding points in the visual field (in degrees) [62]. As previously shown by Gattass et al., cortical magnification in cortical area V4 ($M$) is related to eccentricity ($E$) in the visual field by $M = 3.01 E^{-0.9}$ [29]. We used this equation to estimate cortical magnification in V4 and applied it to convert probe separation distances to their corresponding cortical distances.

Separation distance for a pair of probes was calculated as their Euclidean distance on the grid in degrees of visual angle. The retinal eccentricity of a probe was calculated as the Euclidean distance between its position on the grid and the position of the fixation point (in degrees of visual angle). For every pair of probes, the magnification factor was calculated using the above equation by Gattass et al., where the eccentricity was calculated as the mean eccentricity of the pair of probes (we call it probe pair eccentricity). The magnification factor value for each pair was then multiplied by their separation distance to compute the cortical distance in mm.

We visualized the discrimination performance as a function of cortical distance. First, we divided cortical distance values into 0.5 mm bins. Then we averaged discrimination performances of pairs whose cortical distance values was fell in the range of certain bin. Similarly, discrimination performance was presented as a function of probe pair eccentricities for probe pairs of the same separation distance. The values of probe pairs eccentricities were divided into 3° bins and the discrimination performances were averaged for pairs whose eccentricity fell in the range of each bin. We also calculated the Pearson correlation between the eccentricity and performance to show the significance of these relationships.

*2.8 Selecting similar receptive fields*

From the observation of the receptive fields, we found that several electrodes responded similarly to stimuli at different positions. These electrodes had receptive fields with similar shape, location, and diameter. Therefore, we applied k-means clustering to group these similar receptive fields. The k-means algorithm clusters $n$ samples into $k$ disjoint clusters $C$, each described by the mean $\mu_j$ of the samples in the cluster, called the centroid. The centroids are chosen to minimize the within-cluster sum of squared criterion:

$$\sum_{i=0}^{n} \min_{\mu_j \in C} \left( \|x_i - \mu_j\|^2 \right) \quad (3)$$

The algorithm has three steps: first, it initializes the centroids (e.g. by choosing $k$ samples from the dataset). The second step assigns each sample to its nearest centroid, and the third step is to update the location of the centroid by taking the mean value of all the samples assigned to its previous location. The algorithm repeats the second and third steps until the difference between the old and the updated location of the centroid is less than a threshold. In this implementation we used the k-means++ initialization in scikit-learn to initialize the centroids to be distant from each other, preventing convergence to a local minimum [63].

When similar receptive fields were identified, we selected 1-3 representative electrodes from each group and repeated the discrimination analysis using the total representative electrodes obtained from all the groups.

*2.9 Influence of noise correlations on spatial discrimination*

Most of the recording sites have correlated trial-to-trial variability in response to the same stimulus (called noise correlation). To determine whether these correlations have an impact on discrimination performance, we repeated the analyses above but after destroying the noise correlations in the training set. As before, the SVM cross validation score was computed on the left-out raw unshuffled data (one-fold), but for each electrode site, the order of trials with the same stimulus in the training set were shuffled and SVM was trained on a correlation-free training set. This procedure destroyed the correlation between pairs of recording sites without changing the response statistics at site. Training on a shuffled dataset makes the decoder blind to noise correlations. Change in the discrimination performance then reflects the contribution of noise correlation to subsequent computations.

We calculated the noise correlation values as the Pearson correlation coefficient of the responses of two electrodes to repeated presentation of a particular stimulus [64]. This value is -1 for perfect negative correlation, +1 is perfect positive correlation, and 0 for no correlation. For each probe position, we standardized (z-score) the responses of each electrode independently and then calculated the noise correlation. We removed trials on which the response of either electrode was >3SD different from its mean [65]. We calculated noise correlation for all the electrode pairs and for all the probes included in the analysis.

## 3. Results

In this section, we first characterized the precision of spatial discrimination in area V4 by analyzing LFPs and MUA recorded thorough a 96-channel multielectrode array implanted into behaving macaque monkeys. Then, we proposed two procedures to significantly reduce the number of electrodes for discrimination. Next, we explained the decoding strategy used by the decoder to discriminate positions at smaller and larger distances. Later, we investigated the capability of band-passed LFPs in discriminating positions. Finally, we identified the structure of noise correlations as a source of information about spatial position.

*3.1 Preliminary analysis*

Neural responses were studied with white square probes presented for 25 ms at random locations on a 10×10 grid while the monkey maintained its fixation on the red point (Figure 1(a); Methods). Figure 1(c) left shows five example probes presented at different positions on the grid. Presentation of each probe started at time 0 and lasted for 25 ms. The trial-averaged MUA and LFP signals from a group of selected 28 electrodes, associated to each probe position illustrated on the left, are shown in Figure 1(c), middle. The traces of neural activity are plotted from 100 ms before until 300 ms after stimulus onset. The red traces represent the MUA spike counts obtained over 25 ms time windows. The blue traces are the corresponding LFP signals. The modulation of neural activity is observable in a period of 50 to 125 ms after stimulus onset for both the MUA and LFP traces. MUA modulation appeared in the form of an increase in the spike counts, while modulation in the LFP occurred in the form of biphasic or triphasic fluctuations, often with a clear increase in the signal's negative amplitude. The modulation intensities depended on the position of the stimulus on the grid and differed across recording sites.

We calculated z-scored evoked response values over narrow, medium, and wide windows separately, and plotted the averaged responses to distinct probe positions. These values were then color-coded separately and displayed at the location of the recording electrode within the electrode array. We then compared the ensemble responses defined over each time window (Figure 1(c), right). From the graphs, the ensemble responses to different probe positions are distinguishable over the recording sites. In addition, more electrodes on the array respond similarly to a probe position for LFP than MUA, which means that the LFP responses involve larger cortical regions than the MUA. This suggests that more inter-electrode correlations may exist in LFP responses. Comparing the responses obtained over the three windows shows significant difference to some probe positions, meaning that they may not carry the same amount of information about spatial positions. In the next subsection we compared the precision of spatial discrimination, over these three response windows, for the LFPs and MUA.

The spatial receptive field map of each recording site was determined over the probe grid by averaging the z-scored responses over trials with the same probe positions. The MUA and LFP receptive fields of the same 28 selected electrodes are presented in red and blue colors respectively in Figure 1(b). These contours are ellipses of full width at half height of the fitted Gaussian receptive fields (see

Methods). We determined the receptive fields three times according to the response calculated over each time window. The receptive fields of electrodes are extensively overlapping and vary in size. With narrower time windows, the receptive fields of MUA were not well predicted, while for the medium and wide windows better localization was obtained. Moreover, the LFP and MUA receptive fields were mostly correlated except for a number of MUA receptive fields that appeared around the upper left area of the grid. Further analysis showed that these receptive fields had small response peaks. We determined the receptive field diameter as the mean diameter of the receptive field ellipses. Using the medium time window, the estimated MUA receptive field diameter (the mean of x-diameter and y-diameter of an ellipse) ranged from 4.64º to 29.53º, while for the LFPs it ranged from 12.06º to 31.73º. Using the wide time window, it changed to range 4.64º to 36.30º for MUA and 6.15º to 35.00º for the LFP. These observations suggest that as individual stimulus is presented across the grid, it evokes reliable responses in multiple recording sites, at each spatial position. This distributed response is a potentially rich source of information for position judgements but should be measured over a right window.

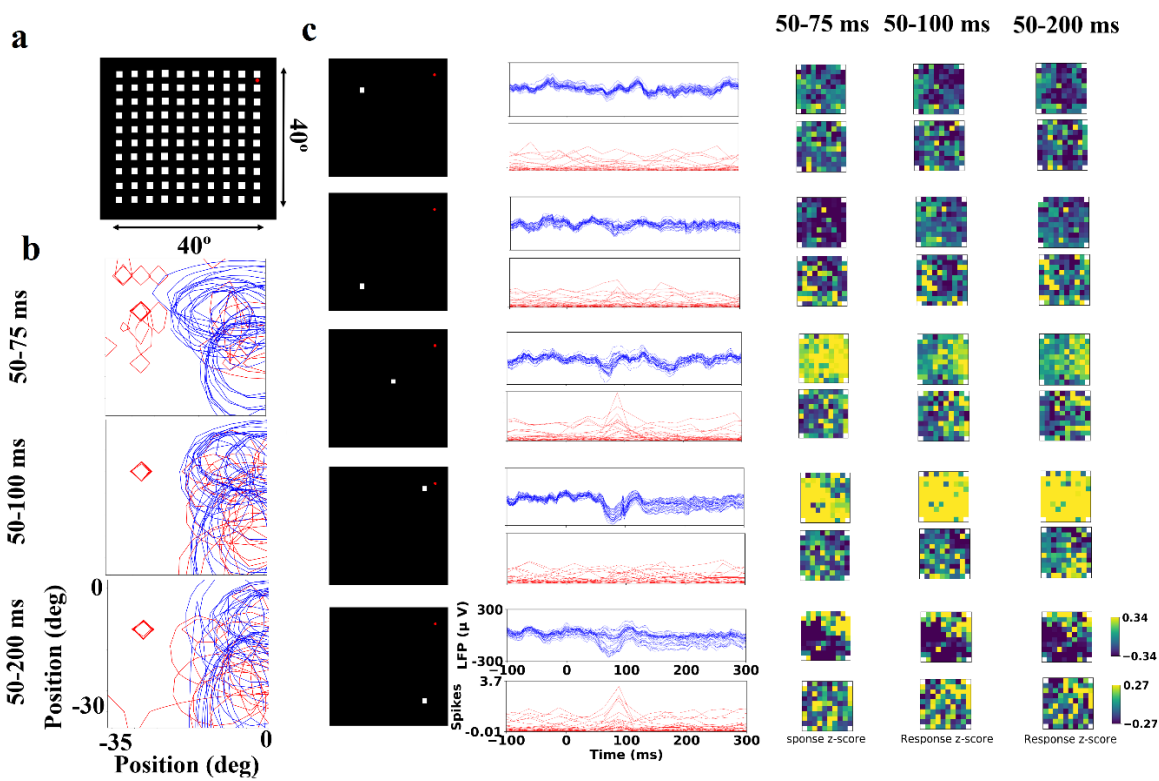

**Figure 1: Responses of neural ensemble to different stimulus positions in visual cortical area V4**. A: All possible visual probe locations on the 10×10 grid spanning 40º×40º of visual space. The red dot is the fixation point (fovea center). B: Estimated receptive fields of a selected group of 28 recording sites in V4 obtained during fixation on the red dot. Receptive field contours are full width at half height of two-dimensional estimated Gaussian; MUA (red) and LFP (blue). Receptive fields were determined based on the trial-averaged evoked responses calculated over three different time windows as presented on their left. The medium size and wider windows show better receptive field localization. The positions in the visual degree are presented on the graph with the red dot (fovea center) as the reference. C: (left) Presentation of visual stimuli at five different positions on the grid and the position of the fixation target at the upper right. (middle) time course of MUA and LFP activity from multiple recording sites (the same selected group of 28 electrodes in B) corresponding to the probe position on their left. For each stimulus, the curves in blue represent the LFP activity and the curves in red show the corresponding MUA spike count sampled over 25 ms time windows. Each trace is the trial-averaged activity of an electrode over trials with the same probe positions. The traces of activity are illustrated between 100 ms before to 300 ms after the stimulus onset. MUA response to stimulus appeared as increase in the spike count while LFP modulation appeared in the form of increase in the negative amplitude of the signal. This modulation for both MUA and LFP appeared around 50-125 ms after the stimulus onset. (right) Responses of electrodes on the array to probe positions illustrated on the left. For the LFP, since the responses mostly appeared as the negative values, we changed the polarity of responses to present more negative values as the higher responses. For each electrode, the (z-scored) responses are averaged over the trials with the same probe position. The responses are calculated for the narrow (50-75 ms), medium (50-100 ms), and wide (50-200 ms) time windows separately. Values on the array are color-coded between -0.27 to 0.27 for MUA and -0.34 to 0.34 for the LFP. The blue color corresponds to small response values and yellow color represent higher values. The white cells on the array show electrically grounded sites.

*3.2 Spatial precision of MUA and LFP in area V4*

In this subsection we characterized the spatial precision of the MUA and LFP responses in V4. To have an unbiased sample, for each time window independently, we excluded electrodes that were not informative about the stimulus positions on the grid (see Methods). This left 28 electrodes for MUA and 91 electrodes for the LFP, when we used the wide response window. For the medium window, for MUA, 29 electrodes were left, and for the LFP, 87 electrodes. Using the narrow window, left 4 electrodes for MUA and 82 electrodes for the LFP.

We separately trained linear SVMs and Radial Basis Function (RBF) SVMs on MUA and LFP response datasets to discriminate differences in spatial position of the probes presented on the grid. For each response window then, we obtained a separate set of probes. Since the receptive fields were mostly located in the right and upper area of the grid, most of these probes were placed in these closer eccentricities. Application of the linear and RBF SVMs showed that both MUA and LFP responses were capable of discriminating the positions of individual probes (Figure 2). The results showed that discrimination performance (cross validation score) depended on the separation distances and eccentricities of the probe pairs.

As the probes were not all placed at the same eccentricity, discriminating probe positions with the same separation distances on the grid do not give the same level of performance. Therefore, we used probes eccentricities and converted their distances to distances on the cortex (see Methods). The discrimination performances of pairs of probes versus their cortical distances were presented in Figure 2(a), (c) for each window separately ((a) applying linear SVM and (c) the RBF SVM). Using the selected informative electrodes for each response window, in all cases, performance increased with cortical distances. Applying linear SVM, with 0.75 mm cortical distance, the average MUA performance for the narrow, medium, and wide windows were 64.7%, 65.2%, and 67.5%, which increased to 73.4%, 84.1%, and 93.8% for 3.75 mm distance respectively (Figure 2(a), left). For LFP with 0.75 mm cortical distance, for narrow, medium, and wide windows, the discrimination performances were 56.5%, 60.8%, and 59.6% that reached 81.5%, 87.0%, and 87.0% for 3.75 mm distance (Figure 2(a), right). These results were predictive for eccentricities and separation distances that were not used in the experiment. Assuming two probe positions with a 1° separation distance at 1° eccentricity, their V4 cortical distance (using $M = 3.01E^{-0.9}$) will be 3.01 mm. At this cortical distance, the MUA discrimination performance with a wide window is around 92% and for LFP it is around 82%.

We performed additional analyses to determine the effect of eccentricity on discrimination performance for probes with the same separation distances. We selected probe pairs with 4° and 8° separations and plotted their performance versus their eccentricity. Figure 2(b) shows the results for each distance for the MUA and LFP and for all three response windows. The results demonstrated that eccentricity plays an important role in spatial resolution: the performance was higher near the fovea and decreased with increasing eccentricity. Exceptions occurred ($p>0.05$) when we used MUA with medium and narrow windows to discriminate 4° separation and a narrow window to discriminate 8° separation. This indicated that the content of information in time windows shorter than 50 ms for MUA is not enough to discriminate between positions. Using the wide window, with MUA, at 4.5° eccentricity and for 4° separation, the performance was 83.1%, while using medium and wide windows with LFP, this performance was 82.6% and 76.8% respectively. For 8° separation, at the same eccentricity, performance with both MUA and LFP were much higher: 95.5% for MUA with the wide window and 85.7% and 95% for LFP with medium and wide windows respectively.

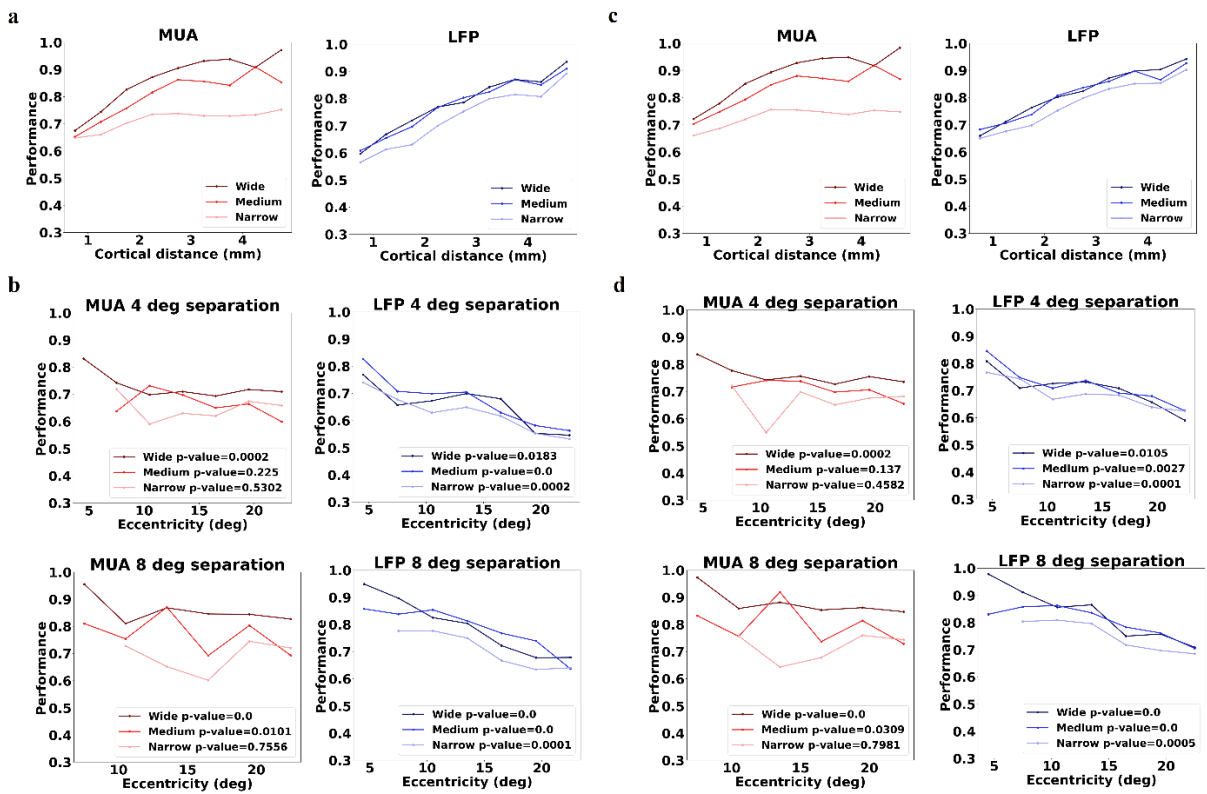

**Figure 2: V4 ensemble response allows spatial discrimination.** Darker colors correspond to the wider response windows. The performances are validation scores of 5-fold cross validation measured as the area under the ROC curve. A: Discrimination performance versus cortical distance (in mm) for MUA (left) and LFP (right) based on responses defined on wide, medium, and narrow windows. The cortical distance values are divided into 0.5 mm bins and the discrimination performances of pairs whose cortical distances fell into the ranges of each bin were averaged. B: Discrimination performances of pairs with 4° and 8° separation distances as a function of eccentricity (in deg) of the pairs for both MUA (left) and LFP (right) for responses defined over wide, medium, and narrow windows. Significance (p-values) of correlation between performance and eccentricity for each response window is presented in the plot. Eccentricity values are divided into 3° bins and the discrimination performances of pairs whose eccentricity fell into the ranges of each bin were averaged. C: Same analysis as section A, applying RBF SVM. D: Same analysis as section B, applying RBF SVM.

We repeated the above analysis with a RBF SVM (Figure 2(c), (d)). The results of using the RBF SVM were similar to the linear SVM, with higher performance (1-4% higher). Although the RBF

SVM yields slightly better discrimination performance, it does not directly assign weights to the responses of the recording sites. Instead it transforms the data into a new space and classifies the transformed data in the new space. In contrast, linear SVM assigns linear weights to each electrode in the discrimination of each pair of positions, which provides more interpretability. Furthermore, linear SVM is faster than RBF SVM, which makes it more practical when the number of discriminations is high. Taken the above factors into considerations, subsequent analyses were performed using linear SVM.

Comparing the discrimination performance of response windows showed that for both MUA and LFP, narrow time windows led to lower performance, while medium and wide windows resulted in much better performances. These results suggest that capturing the content of information about the spatial precision requires a response window of at least 50 ms. For MUA, the wide window obtained better performance over the medium window. For LFP, medium and wide windows had very close spatial precision with a mean difference of only 0.71% for the wide versus the medium window. This suggests that the information content for the spatial discrimination is mainly accumulated between 50-100 ms after the stimulus onset. This property makes the medium size window more practical for the use of LFP when temporal resolution is important. Therefore, for the subsequent analysis we used medium response window for LFPs and the wide window for MUA.

In practical applications, we are interested in smaller probe separations with high spatial precision. As we showed above, if we present the probes in smaller eccentricities, the discrimination performance will be much higher, as a much larger area of the visual cortex is dedicated to the fovea representation. Therefore, if we implant the same array in the foveal region of V4, it will yield far better resolution. In the next subsection, we presented new methods to select a smaller population of electrodes systematically while increasing the performance for a particular value of separation distance.

*3.3 Minimizing the number of electrodes*

In this subsection we proposed and compared two methods to reduce the number of electrodes for use in the discrimination, without significantly impairing the performance. In the first method, we clustered electrodes whose receptive fields are similar while in the second method we used linear SVM weights to select minimum number of the best electrodes.

*3.3.1    Clustering electrodes with similar receptive fields*

Comparing the receptive fields of individual electrodes showed some level of similarity in their shapes and locations that made it possible to group them into a few classes. Using k-means clustering to group the receptive fields of the informative electrodes, we found that 3 clusters gave the best results for both MUA and LFP. Figure 3(b), bottom, shows the receptive fields that fell into the same group. To select a smaller population for discrimination, from each group, we selected 1-3 representative electrodes with the highest receptive field peaks and repeated the above discrimination analysis. Figure 3(a) compares the discrimination performance as a function of cortical distance for representative and all the informative electrodes. The performances were obtained with the 8 representative electrodes for MUA and 9 for LFP. Using the representative electrodes, the average MUA performance decreased by 3.2% while for LFP it decreased by 3.9%. These results suggest that performance was marginally affected by removing redundant electrodes from the analysis.

Next, we presented the discrimination performance of MUA and LFP as a function of eccentricity for 4° and 8° separations (Figure 3(b)). With 4° separation and <8° eccentricities, for MUA, using the representative electrodes, the performance was increased by 6% compared having all the electrodes. For LFP, this change in performance was a small decrease of 0.8%. For the 8° separation, the average performance of MUA and LFP decreased by 4.6% and 4.7% respectively. Moreover, in all cases, the correlation between performance and eccentricity were significant ($p<0.05$). Clustering of similar electrodes, then can help to reduce the number of electrodes to less than 10 while maintaining reasonable spatial precision.

a
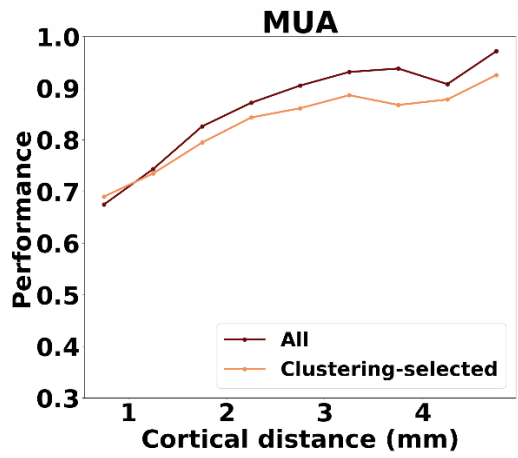
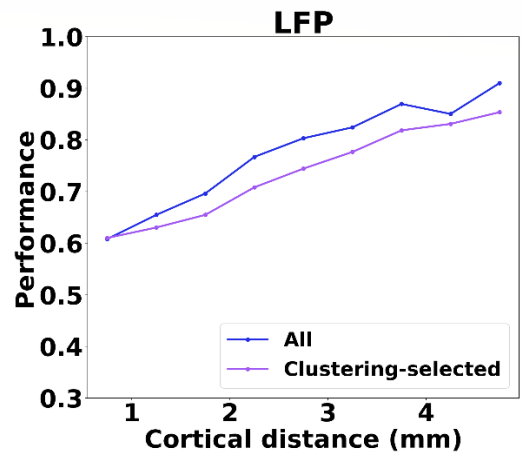

b
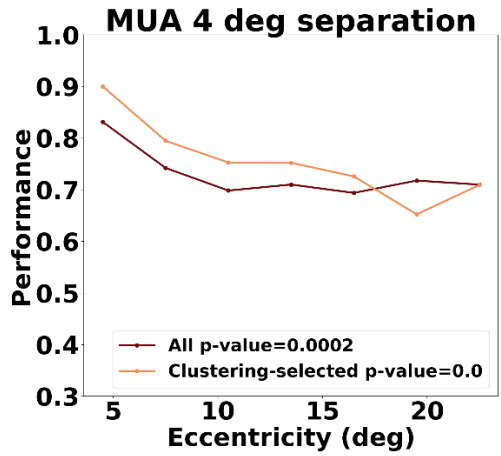
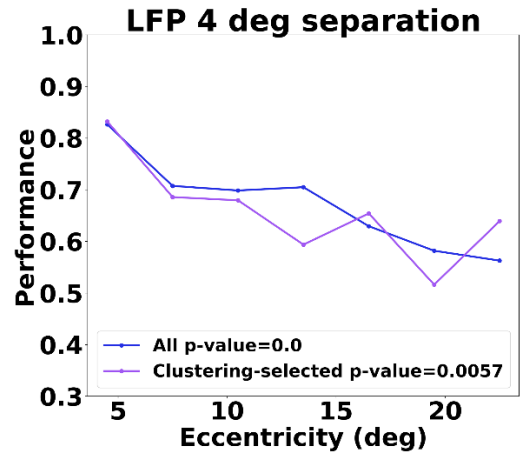

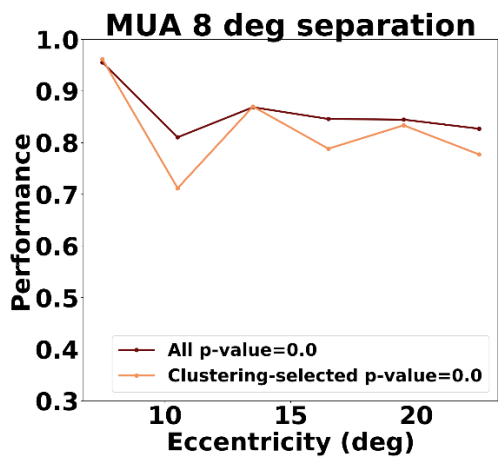
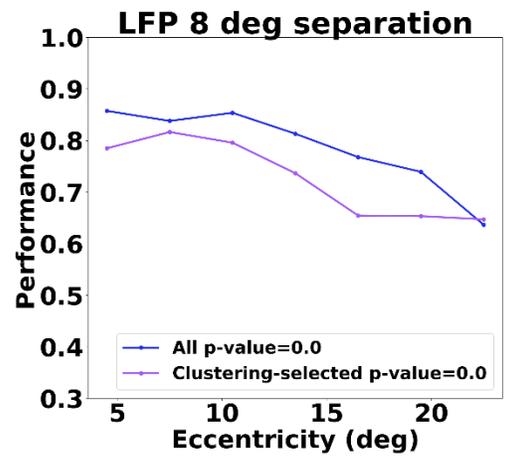

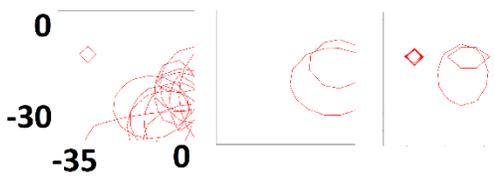
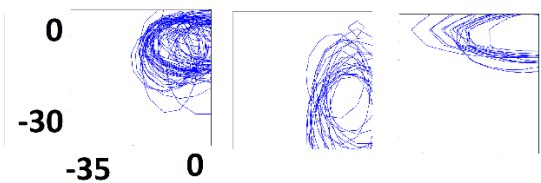

**Figure 3: Spatial discrimination with a smaller subset of electrodes selected using clustering.** Similar receptive fields were clustered into 3 groups and 1-3 representative electrodes with highest receptive field peaks were selected from each group for use in the discrimination analysis. Three (k=3) groups for MUA (left) and three for LFP (right) were obtained. Dark red and blue colors represent for MUA and LFP respectively, when all the informative electrodes were included in the analysis. Orange and purple colors show MUA and LFP when only representative electrodes were included in the analysis. A: Comparing spatial discrimination as a function of cortical distance when all the informative electrodes were used and when only representative electrodes were used. The cortical distance values are divided into 0.5 mm bins and the discrimination performances of pairs whose cortical distances fell into the ranges of each bin were averaged. B: Comparing spatial discrimination versus eccentricity for probes with 4° and 8° separations. Significance of Pearson correlation (p-values) are presented on the plots for each curve. Eccentricity values are divided into 3° bins and the discrimination performances of pairs whose eccentricity fell into the ranges of each bin were averaged. (bottom) The contours of similar receptive fields over the grid coordinate system with fovea location as the reference (red for MUA and blue for the LFPs).

*3.3.2    Selecting electrodes with the highest importance values*

We sorted the electrodes based on their overall importance and selected the 4, 6, 8, 10, and 12 most important (best) electrodes, and for each set separately we performed the discrimination analysis as above. Figure 4(a) illustrates discrimination performance versus cortical distance for each set of the selected electrodes, as compared to the performance when the full set of the responsive electrodes were used. For both MUA and LFP responses, the full set of electrodes (black curves), gave the highest performance over all the discriminations. This was specifically obvious when the cortical distances were larger. Although, in shorter cortical distances, 6-12 electrodes obtained slightly better performance than all the electrodes, this difference was negligible over all the discriminations. As before we also illustrated discrimination performance versus eccentricity for 4° and 8° separations (Figure 4(b)). For the MUA, for 4° separation in closer eccentricities, with 4 and 6 electrodes, performance was 2% and 0.1% higher than with the full set of electrodes respectively. For LFP also, 0.7% higher performance was obtained with 8 electrodes, for 4° separation in the closer eccentricities. In both MUA and LFP, for 4° separation and closer eccentricities, 10 and 12 electrodes gave lower performance. Overall, in all cases, the performance differences did not exceed 10%. For 8° separations however, the full set of electrodes gave better results than the each of the sets. Comparing the overall performance of MUA and LFP, we concluded that the minimum number of electrodes for fine discriminations (4°) without impairing performance (and also achieving $p<0.05$) for LFP was 8, while for MUA was 4.

Although the above results showed that systematically selecting electrodes could slightly improve the discrimination performance and at the same time significantly reduce the number of electrodes, the selection strategy includes all the discriminations with <15° separations. As the goal of cortical visual prosthetic devices is to discriminate stimuli of small separations (high spatial resolution), and in closer eccentricities, we repeated the above analysis after averaging the importance values only over the discriminations with 4° separation at closer than 8° eccentricities. We sorted the electrodes based on these new importance values and again selected the best 4, 6, 8, 10, and 12. We repeated the discrimination analysis and illustrated the performance as the function of eccentricity for the 4° separation (Figure 4(c)). For MUA, at 4.5° eccentricities, the performance was always around 10% higher than the performance obtained with all the electrodes. For LFP, at 4.5° eccentricities, using 8, 10, and 12 electrodes, performance reached 86.4%, 88.2%, and 88.4% which were 3.8%, 5.6%, and 5.8% higher than the performance obtained with all the electrodes. Further analysis (not shown) on LFP with more than 12 electrodes showed that performance reached its maximum value (90%) with 16 electrodes. To obtain insight about the distribution of these best electrodes on the cortex, we plotted the location of the 6 best electrodes for MUA and 16 best electrodes for LFP (Figure 4(c), bottom of the curves). According to the cortical magnification factor in V4 [29], 4° separation at 4.5° eccentricity subtends a distance of 3.11 mm on the cortex. This amount of distance in a square (3.11×3.11 mm$^2$) can cover 64 electrodes of the Utah array. However, the results of this analysis showed that precise discrimination can be made with a much sparser distribution of electrodes, which in turn produces less damage to the cortex.

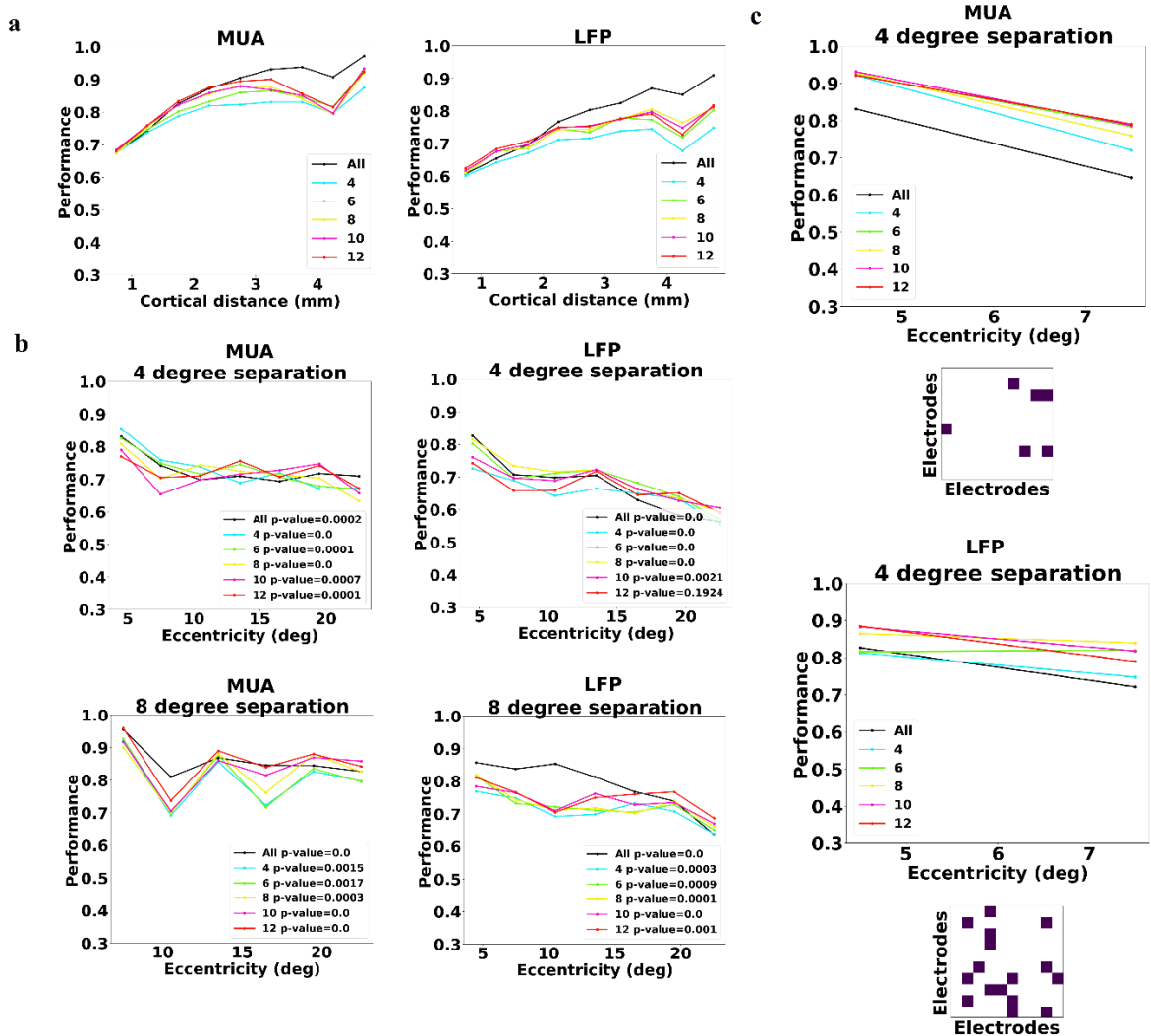

**Figure 4: Discrimination analysis with the best (most important) 4, 6, 8, 10, and 12 electrodes in blue, green, yellow, purple, and red colors respectively.** Discrimination with all the electrodes showed in black. A, B: importance of each electrode was calculated by averaging its importance values over all the discriminations with separations <15°. A: Discrimination performance versus cortical distance (in millimeters) for MUA (left) and LFP (right) when the best electrodes were used. The cortical distance values are divided into 0.5 mm bins and the discrimination performances of pairs whose cortical distances fell into the ranges of each bin were averaged. B: Discrimination of probes with 4° (top) and 8° (bottom) separation versus the mean eccentricity of pairs in visual degrees using the same sets of the best electrodes as A. MUA results are presented on the left and LFP on the right.

Pearson correlation significance (p-values) are presented on each plot associated to the results of each electrode set. Eccentricity values are divided into 3° bins and the discrimination performances of pairs whose eccentricity fell into the ranges of each bin were averaged. C: Selecting the best electrodes was based on new importance values obtained by averaging importance values of discriminations with 4° probe separation and <8° pair eccentricity. The electrodes were sorted by these new importance values. As only the closer eccentricities were important, the plots are illustrated for smaller range of eccentricities. At the bottom of each plot the locations of the best electrodes are presented on the electrode array. These best 6 electrodes for MUA and 16 electrodes for LFP can achieve better performance than using all the electrodes for 4° probe separations.

We used two different methods to minimize the number of the electrodes for discrimination of spatial positions. Comparing the two methods showed that selecting electrodes based on their importance values provided a more systematic strategy which at the same time minimized the number of electrodes and increased the performance for both MUA and LFP. For the subsequent analysis in which we need to select the best electrodes, we used their importance values.

*3.4 Coding strategies*

We next analyzed the coding strategy for position discrimination by examining the weights recovered by the SVM. This analysis shows that how the weights assigned to the electrodes can be used to discriminate a position from other positions at a specific distance and how these values are related to the spatial tuning of the underlying electrodes. To estimate this information, the weights were first normalized for each discrimination. For each probe position included in the analysis, we selected discriminations made between that probe and the other probes located at a particular separation distance from it and averaged the weights for these discriminations. We then presented these values for all the probes at their location on the grid.

We investigated the coding strategy for 4°, 8°, and 12° separations for MUA and LFP independently. Examining individual electrodes suggested that, for small separations (4°) the peak of the receptive fields of more informative sites was offset from the target, such that the target position to be discriminated was situated on the flanks of the receptive field (Figure 5). In contrast, for large position separations (12°), sites were more likely to be informative if one of the targets was near the centre of the receptive fields. We observed a transition in the distribution of the weights between the two coding schemes over a range of target separations (8°). This pattern was observed for both MUA and LFP. However, since LFP responses appeared in the negative amplitude of the signal, negative weight values followed this pattern (Figure 5(b), right). These results are in agreement with the coding schemes for spatial position discrimination previously reported in area MT [27].

There was no significant correlation between the weights assigned to a particular electrode and the cortical distances of the included pairs of probes indicating no association between weights and the cortical distance both for MUA and LFP ($p>0.05$). The same analysis on the averaged importance values of the four best electrodes and the cortical distances showed similar results.

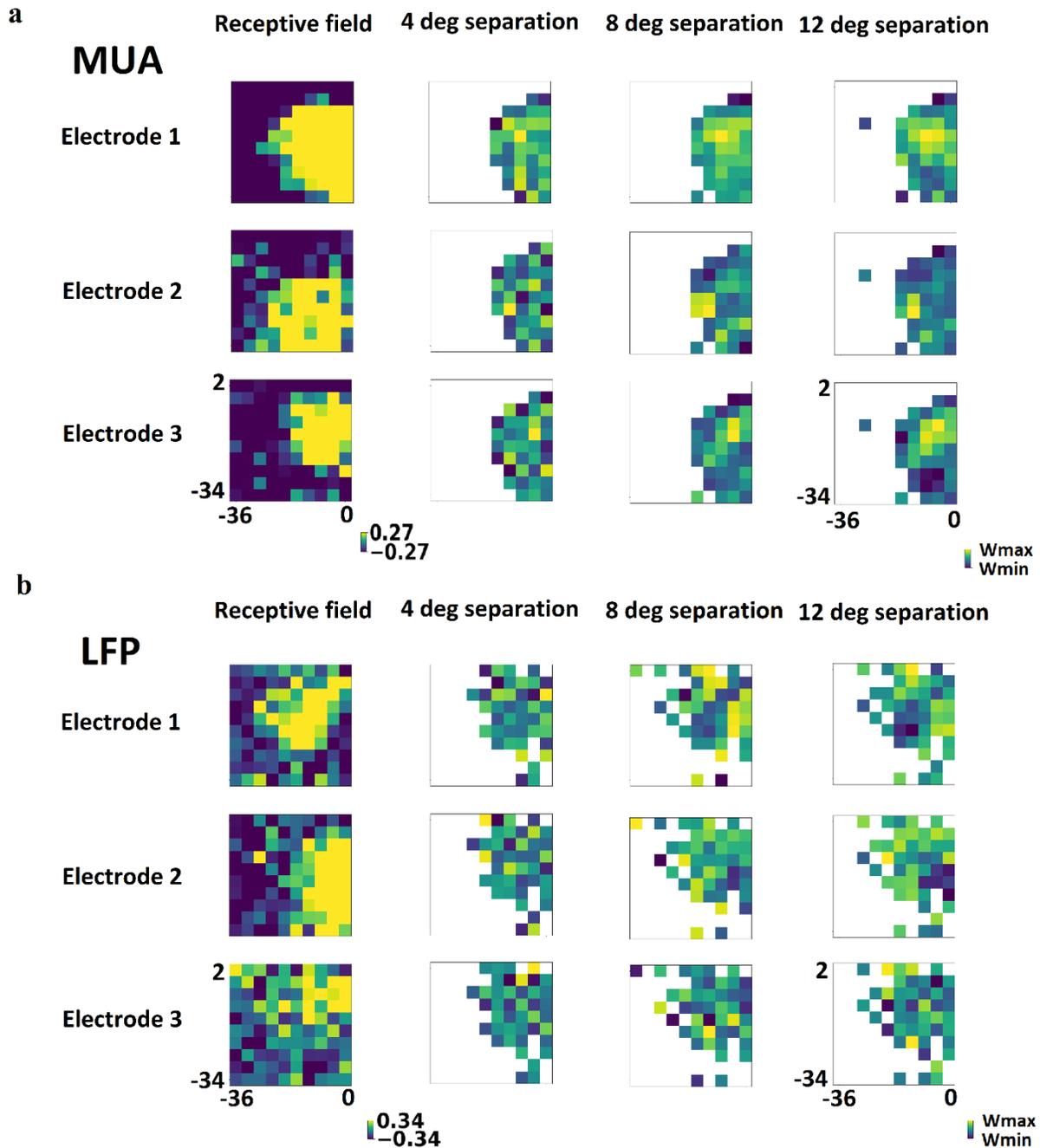

**Figure 5: The coding strategy of individual sites for position discrimination of small and large separations.** The degree of color lightness indicates polarity of values; lighter colors show more positive and darker blue colors show more negative values. Spatial receptive field profile is shown for each electrode on the left. The spatial distribution of the weights assigned to each site for discrimination of a probe from the other probes with a fixed distance from it. This value is calculated for each probe and is presented at its location on the grid. The change in the coding strategy can be observed from 4° to 12° separation in the visual field (left to right). In smaller separations, most of the target positions are at the flanks of the receptive fields while for larger separations they are close to the center of the receptive fields. The values of the discrimination weights are visualized over the ranges between their maximum and minimum to capture the weight variations specific to each separation distance. A: MUA results: The range of response values for the receptive fields are set to -0.27 to 0.27. The best three

electrodes for MUA are selected for this analysis. As MUA responds by positive increase in the spike count, from 4° to 12° separation, the more yellow regions move from the flank of receptive fields to the centre of receptive fields. B: LFP results. The range of response values for the receptive fields are set to -0.34 to 0.34. The best three electrodes for LFP are selected for this analysis. As the LFP respond by negative peaks, the darker colors are the representative of targets in the weight distribution. As the results show, from 4° to 12° separation, the darker regions move from the flank of receptive fields to the centre of receptive fields.

*3.5 Spatial precision of LFP at different frequency bands*

We next sought to determine whether specific LFP frequency bands carry different amounts of information about stimulus positions. For each band-limited LFP, we calculated the responses using the medium time window. We used the same full set of responsive electrodes, as used for the broadband LFP with the medium window, and repeated the SVM discrimination analysis.

Figure 6 compared the discrimination performances obtained for each frequency band. Plotting performance versus cortical distance (Figure 6(a)) showed that for no frequency band did the performances exceed 70%. This suggests that the information about stimulus position is distributed across frequencies. To study the relationship between band limited performance and eccentricity, we repeated the above analysis for 4° and 8° separations (Figure 6(b)). Although there was a correlation between performance and eccentricity, with 4° separation using theta frequency bands and with 8° separations using the gamma frequency band, no overall significant relation to eccentricity was found as a source of information about spatial position. These findings indicated the necessity of using broadband LFP in extracting information about the spatial positions.

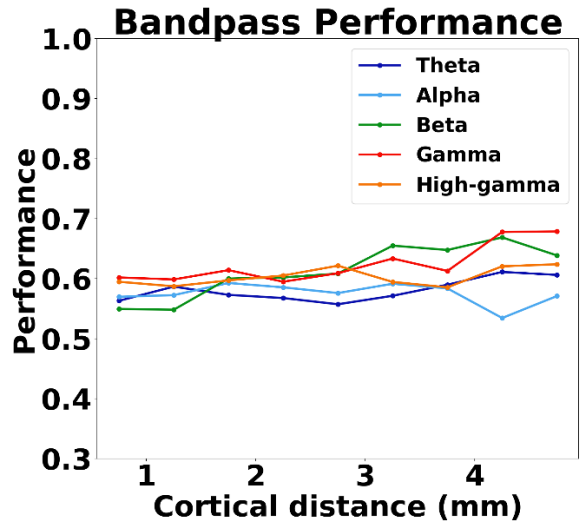

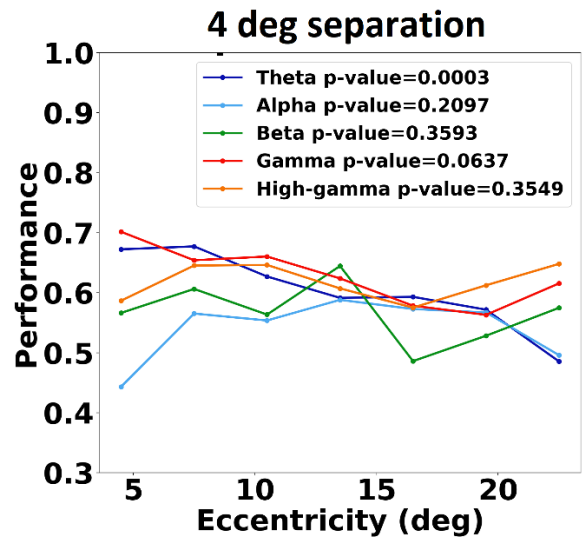

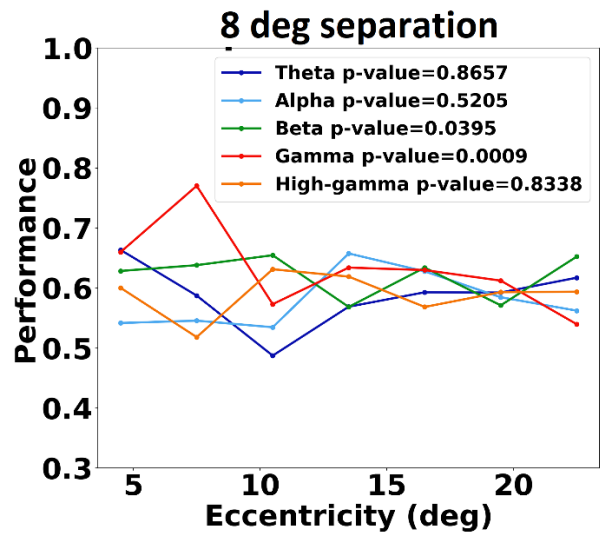

**Figure 6: Discrimination performance of band passed LFPs.** Responses for each band passed signal were calculated using the medium window. A: The band passed LFP performance versus cortical distance for theta (4-8 Hz), alpha (8-12 Hz), beta (12-30 Hz), gamma (30-50 Hz), and high gamma (50-80 Hz) frequency bands. The cortical distance values are divided into 0.5 mm bins and the discrimination performances of pairs whose cortical distances fell into the ranges of each bin were averaged. B: Discrimination performances of the same band passed LFPs (as A) as a function of eccentricity for 4° and 8° separations. The significance (p-values) of each band is presented on the plots. Eccentricity values are divided into 3° bins and the discrimination performances of pairs whose eccentricity fell into the ranges of each bin were averaged.

*3.6 Influence of noise correlations on spatial discriminations*

The receptive fields in V4 are highly overlapping, and therefore the neurons are likely to receive input from similar groups of afferent neurons. Thus, some of their responses will be shared with other neurons and consequently, correlated [15, 66]. These interneuronal correlations in the responses might be used by decoding algorithms [54]. Noise correlations, which are the shared trial-to-trial variability in responses, are critical for understanding the representation of stimuli in neural ensembles [67] and have been shown to be related to the amount of information in a neuronal population [65].

To assess whether in our analyses, SVM uses noise correlations, we trained it on datasets in which the correlation structure was destroyed and cross validated it on the real unshuffled data (Figure 7(a)). The removal of noise correlations from the training set impaired the ability of the decoder to interpret real neural activity and reduced the decoder performance (Figure 7(b)). Averaging the performance over all the discriminations showed that this reduction was around 3.6% for MUA and 13.6% for the LFP. This reduction was around the same level over all cortical distances. This suggest that the change in the computational performance was independent of cortical distances. We also illustrated the performance as a function of eccentricity for 4° and 8° separations (Figure 7(c)). With MUA, the effect of noise correlation was more pronounced in discriminations with 4° separation with 5.4% reduction compared to 8° separation with 2.6% reduction in the performance. This pattern was different for LFP with 12.5% reduction for 4° separation and 16% reduction for 8° separation. These results indicate the importance of noise correlations in finer discriminations when MUA is used and in coarser discriminations when LFP is used. In addition, it suggests that LFP performance highly depends on the noise correlations. Altogether, findings suggest that the interneuronal correlations contain information that could in principle be extracted by downstream regions.

To study the effect of noise correlation in smaller population of electrodes with comparable results, we selected the best six electrodes over all the discriminations for both MUA and LFP separately. Repeating the analysis for this population showed that the average reduction in the MUA performance was 4.6% and for LFP it was 12.1% (Figure 7(d)). In other word, by reducing the number of electrodes, the effect of noise correlation on performance stayed around the same level. This suggests that using smaller number of electrodes still requires the same considerations about the noise correlation.

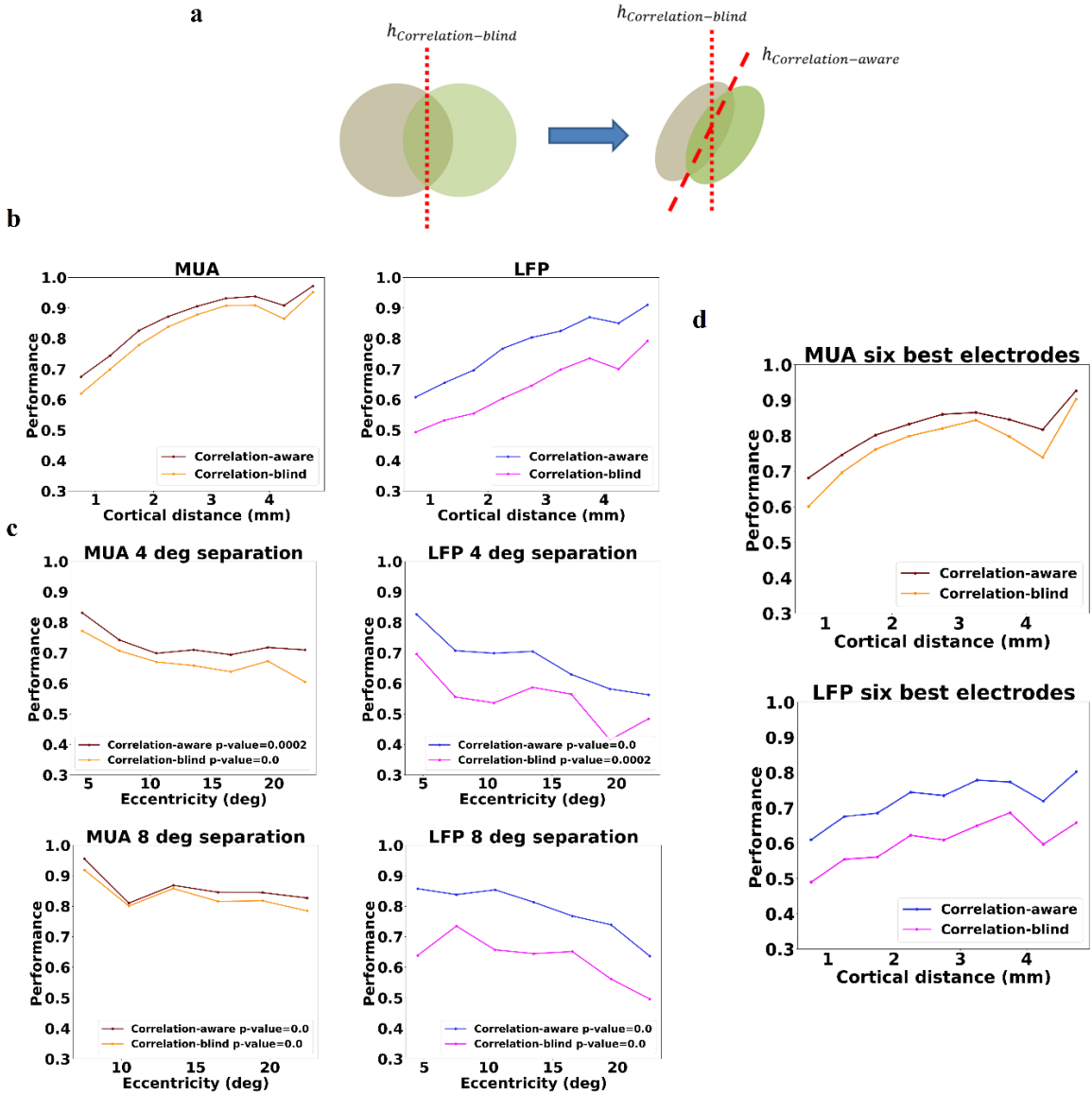

**Figure 7: Effect of noise correlations on spatial discrimination.** Dark red and orange colors represent correlation-aware and correlation-blind performances of MUA respectively, while blue and purple colors are representative of the same measures in the LFP. The correlation-blind performances are the cross validation scores when SVM is trained on the shuffled data. A: (left) Training decoder on shuffled data with noise correlations removed. $h_{Correlation-blind}$ shows the hyperplane optimized on the correlation-blind data. (right) original unshuffled data are discriminated with $h_{Correlation-blind}$ and compared to the discrimination with $h_{Correlation-aware}$ that is optimized on the original unshuffled data. B: Discrimination performance versus cortical distance for correlation-blind versus correlation-aware responses. (left) MUA (right) LFP. The cortical distance values are divided into 0.5 mm bins and the discrimination performances of pairs whose cortical distances fell into the ranges of each bin were averaged. C: Discrimination performance as a function of eccentricity for 4° (top) and 8° (bottom) separations comparing correlation-blind and correlation-aware response data (left) MUA (right) LFP. Pearson correlation p-values are shown on each plot indicating the significance of the relationship. Eccentricity values are divided into 3° bins and the discrimination performances of pairs whose eccentricity fell into the ranges

of each bin were averaged. D: Same as A when the best six electrodes for MUA and the best six electrodes for LFP are selected for the discrimination.

We next measured the noise correlations. To understand the dependence of noise correlation on the stimulus position, we selected the best two electrodes in the discrimination for LFP and MUA separately and calculated their noise correlation for the probes included in the analysis. We then plotted the resulting values at the location of probes on the grid (Figure 8(a)). Comparing the results to the response profiles of the electrodes indicated that the stimulus can induce higher or lower levels of correlation depending on the receptive field location. For both MUA and LFP, the magnitudes of noise correlations were increased when the probe was in the receptive field of both sites or in the receptive field of neither of the sites. This pattern has been observed before for spiking activity [15, 64, 68, 69]. Correlations were relatively low when the stimulus was in the receptive field of one site but was not in the receptive field of the other site. This pattern has also been observed in area MT [27].

To understand how noise correlations change over cortical distances, we calculated the noise correlations for all the electrode pairs included in the analysis and sorted the values by the inter-electrode distances on the cortex. Figure 8(b) shows noise correlation versus inter-electrode distances for six different probes for both MUA and LFP. These six probes were selected from three independent ranges of eccentricities with two probes in each range that were presented with the same colors. The results showed that for MUA, overall noise correlation dropped slightly (<0.1) with increasing the inter-electrode distances, in agreement with [67]. This decline in noise correlation for LFP was much bigger and for some probe locations reduced by >0.2 for 4 mm increase in inter-electrode distances. This level of dependence of LFP on the noise correlation can explain the high drop in discrimination performance of the correlation-blind versus the correlation-aware responses. As we found before, discrimination performance increases with increasing cortical distance. The results here suggest that more distant points on the cortex have lower levels of noise correlation, and thus will be easier for the decoder to discriminate. In addition, since LFP produces highly correlated responses, even in more distant points on the cortex, responses will be more difficult than MUA to discriminate. Success in discrimination with LFP signals then, can be achieved by selecting electrodes optimally and as far as possible on the cortex and presenting stimuli at smaller eccentricities. The curves for probes at different eccentricities also showed no particular relationship between eccentricities and noise correlation. As we can see in the figures, the correlations for probes at the same eccentricity ranges had different levels of noise correlation for both MUA and LFP signals.

We also investigated the relationship between the mean importance of pairs of electrodes and their noise correlations (Figure 8(c)). The results showed that noise correlation drops with increased overall importance of the electrodes, and this drop is more pronounced in LFP. This suggests that electrodes that are important in discrimination of two positions are less correlated. In addition, no significant relationship was observed between the eccentricity of probes and their noise correlation as a function of electrodes importance. As we could select the minimum number of electrodes optimally, we actually selected less correlated electrodes in the discrimination analysis. Therefore, we minimized the redundant information but still kept important information that existed in the noise correlations for the discrimination.

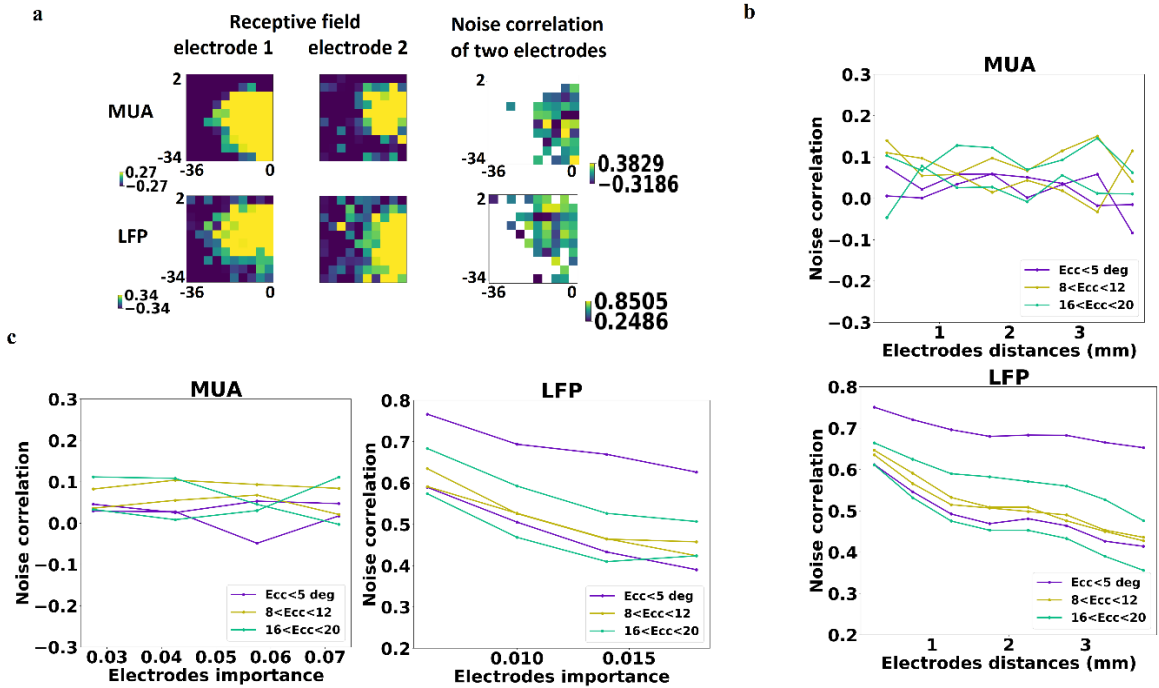

**Figure 8: Dependence of noise correlation on stimulus position, electrodes distances, and electrodes importance.** Noise correlation values are the Pearson correlation coefficients and are between -1 to 1. A: Noise correlation of the best two electrodes (independently for MUA and LFP) in response to different probe positions on the grid. The range of noise correlation values is presented on the grid (on the right) and compared to the receptive fields of the two electrodes (on the left). We limited the range of the receptive field values on the graph between -0.27 to 0.27 for MUA and -0.34 to 0.34 for the LFP. The noise correlation values for both MUA (top) and LFP (bottom) are limited between Mean±1.5SD of the values on the graph. The degree of color lightness indicates polarity of values; lighter colors show more positive and darker blue colors show lower (more negative) values. The positions on the graphs are shown with the fovea center as the reference. B: Noise correlation versus distances of the electrode pairs for responses to six different probe positions placed at three separate eccentricity ranges- with two probes in each range. Purple color for the eccentricities less than 5° from the fovea, brown color for eccentricities between 8° and 12°, and green color for eccentricities between 16° and 20°. (top) MUA (bottom) LFP. C: Noise correlation versus mean importance of the electrode pairs. The importance value of each electrode is obtained as the mean of the squared weights calculated over all the discriminations (with <15° probe pair separations). The same probe positions and color codes as B are used. (left) MUA (right) LFP.

## 4. Discussion

While most of the previous cortical visual prosthesis studies have been performed in V1, extrastriate visual areas offer sampling of a much larger region of visual space with the same coverage of electrode implants. Receptive fields in extrastriate areas are large and have extensive spatial overlap and therefore, spatial precision is unlikely to match the maximum precision in area V1. However, previous analysis have shown that the interaction between multiple receptive fields can provide relatively high precision that is behaviorally relevant [27]. Here, we studied the two dimensional spatial precision of V4 as an intermediate extrastriate region in the visual processing hierarchy. We chronically implanted Utah electrode arrays in area V4 of macaque monkeys and recorded LFP and MUA activity, while individual stimuli at different positions were presented. Discrimination of visual features has been previously shown to rely on ensembles of neurons with broad tuning curves [27, 54, 70]. Our results used a similar representation for the spatial positions in two dimensions. We found that despite wide difference in the receptive fields of the electrodes, both LFP and MUA were able to discriminate spatial positions.

*4.1 Spatial discrimination of MUA and LFP*

We found that both MUA and LFP responses are capable of precisely discriminating spatial positions and that this precision increases with increasing the distance between the corresponding representation of positions on the cortex. The performance levels obtained for MUA and LFP were comparable with that for MUA being an average of 6.7%. The results showed that the discrimination performance was highest at smaller eccentricities $<°8$. Probes that were presented closer to the fovea location then, had more distant positions on the cortex and could be discriminated more easily. Therefore, for prosthetic applications in which high spatial resolution is required, implanting electrodes in the fovea representation on the V4 cortex would give higher performances even for very small spatial separations.

Comparing linear and RBF decoders showed that neural responses in V4 corresponding to spatial positions are mostly linearly separable. SVMs are appropriate when the number of samples is less than the number of features. Here, we had 14-21 samples for each probe while the number of features was a subpopulation of electrodes and most of the time was larger than the number of samples. If the number of times that each probe position presented was more than 50 times we could use even simpler decoders such as a Linear Discriminant analysis (LDA) [71]. An increase in the number of trials then, can provide a better estimate of the linear decision boundary for discriminating spatial positions. For cortical visual prosthesis applications in which a precise use of current stimulation is needed, having an accurate estimation of the decision boundary will be important.

There are two advantages to using LFP instead of MUA: 1) LFPs are more durable: they can often be measured reliably for years after the implantation of the multi-electrode arrays [33, 34]. Therefore, for the prosthesis applications in which both recording and stimulation are important they are more applicable. 2) LFP responses can be captured over the medium window 50-100 ms while for MUA to achieve maximum performance, the wide window 50-200 ms is required. For real-time applications where the system needs to maintain temporal resolution, using the LFP will be more applicable. However, to have good precision with LFPs, implantation of electrode array in the fovea representation is critical.

*4.2 Minimizing the number of electrodes*

One of the main challenges in the development of prosthesis is the damages it causes to the tissue. This damage can result from the insertion and chronic presence of the intracortical electrode [72-75], excess charge delivery [76, 77], chronic electrical stimulation [78-80], or thermal damage [81]. In all these cases, reducing the number of electrodes can reduce the damage to the tissue. This reduction in the electrodes is obtained by selecting V4 instead of V1 as fewer electrodes are needed to cover the same extent of the visual field. For V4 itself, we also showed that we can further reduce the number of required electrodes without impairing performance. We found that the minimum number of electrodes needed to obtain high spatial precision for both MUA and LFP is much lower than all the electrodes on the array (6-10 for MUA and 8-16 for LFP). Although previous literature has shown that using fewer electrodes selected randomly for decoding decreases the performance [27], we showed that if those electrodes are selected systematically, we can obtain higher performance for fine discriminations. For applications in which multiple cortical areas are implanted with electrodes, this could significantly reduce the number of recording (or stimulating) electrodes.

We proposed and compared two methods for selecting the best electrodes: clustering of electrodes with similar tuning and selecting electrodes based on their weights assigned by SVM. There are several advantages of using SVM weights versus clustering analysis: 1) In applications where finding the receptive field map of the electrodes is difficult or impossible, SVM weights can be used. 2) Calculation of performance and weights are performed with the same algorithm, while clustering is a completely separate analysis. 3) Using the weights provides a more systematic procedure, as we can sort the electrodes for an arbitrary set of discriminations. 4) Using the SVM weights yielded better discrimination for both MUA and LFP. 5) Depending on the nature of experiment, k-means clustering analysis may not always give the best grouping of electrodes and more complicated clustering algorithms are required to overcome complexities in the responses. But, SVM weights are computed linearly as part of the discrimination process.

*4.3 Preprocessing steps*

Preparing LFP signals for use in spatial discrimination analysis requires a number of preprocessing steps. The initial and the most important one is to extract the stimulus-modulated responses from the signals. The appropriate response measure is found under some assumptions about the signal and verification of the ability of the response in discriminating fine or coarse differences in the visual features. Defining the response measure is a features extraction step that has been obtained partly through trial-and-error. Such approaches can bias the results, as they entail assumptions on the response measures. For example, in the LFP analysis, we used a time window and defined the responses as the mean amplitude over this window. One assumption is that the information about the stimulus is encoded in the LFP amplitude. The other one is that the relevant responses occur for example between 50-100 ms after stimulus onset. Although using these assumptions provides relatively precise discrimination results, they are not automatically used as a part of the classification algorithm (here SVM). There present a set of techniques called feature learning that can replace manual feature engineering by automatically discovering features and learning them to perform specific tasks [82]. Examples include artificial neural networks. These algorithms can be used to learn response features and classify the LFP signals directly without the need to define the responses manually. However, using these techniques require many training examples (trials) to work accurately.

In our analyses, we investigated the need for filtering the LFP signals to specific frequency bands in order to extract more precise spatial information. Our analysis then showed that the information about the position is encoded in the broadband LFP. The performance thus dropped significantly by limiting LFP to specific frequency bands. This suggests that for spatial discrimination in V4, there would be no need for additional preprocessing step of filtering LFP.

*4.4 Microstimulation pattern to generate a phosphenated percept*

Cortical visual prostheses offer the promise of restoring vision in patients with acquired blindness resulting from damaged eyes or optic nerves [1, 83]. The functionality of these devices relies on the hope to localize multiple phosphenes in the visual field by electrically stimulating multiple implanted electrodes in the visual cortex [4-9]. Most previous studies have used concurrent electrical pulse stimulation of single or multiple electrodes to generate phosphenes and control their characteristics. However, there is no report of restoring a visual scene by localizing multiple phosphenes experimentally. One explanation for these failures is that the neural activity patterns induced by current pulses are not similar to the natural patterns, and so the downstream circuits cannot easily interpret them. In contrast to visual stimuli that activate neurons selectively, electrical stimulation activates random set of neurons in a close vicinity of the stimulating electrodes [84]. Electrical stimulation of early visual areas (V1/V2/V3) can generate simple percepts, but this non-selective activation of neurons may not effectively propagate to higher visual areas to produce natural vision [85]. This explains the observation that electrical stimulation of late visual areas (e.g. IT) sometimes fails to induces a percept [86], as they require generation of complex activation patterns. This also explains the need for extensive training sessions before patients are able to detect percepts induced by electrical stimulation in both early and higher visual areas [86]. With training, as a result of brain plasticity in the adult brain, new activity patterns can be learned and electrical stimulation of any part of the cortex can be detectable [85]. However, the induced perception is still distinct from the natural activation patterns.

Imitating natural activity patterns generated in response to phosphene-like stimuli, provides a systematic way to control the positions of multiple phosphenes. Since those activity patterns are familiar to the cortex, they may easily become detectable and be interpreted by downstream circuits. To imitate a natural activity pattern, a novel stimulation paradigm, called dynamic current steering, was suggested [87]. However, this paradigm still doesn't imitate natural responses but instead was *inspired* by the V1 responses.

For localization of multiple percepts at different positions, their corresponding imitated activity patterns should be discriminable. As our results showed, spatial positions are discriminable using both MUA and LFP responses in V4. The precision of this discrimination depends on the distance between their representations on the cortex. As the magnification factor increases with decreases in eccentricity, the positions located at smaller eccentricities will be placed at farther distances on the cortex, and thus can be discriminated more precisely. By implanting electrodes in the foveal representation of V4, we can precisely discriminate probes that are located at very small distances. As we calculated earlier in the results section, probes with 1° separation at 1° eccentricity are located with 3.01 mm distance from each other on the cortex. At this cortical distance, 7-8 electrodes of Utah electrode array, similar to what was used in this study, can be embedded. We conclude that for prosthetic applications, electrodes should be implanted in the foveal representation and the spatiotemporal patterns of electrical stimulations should be able to reconstruct discriminable natural responses to stimuli at different positions.

To generate such patterns of electrical stimulation, the stimulation parameters need to be tuned. As we found in our results, discrimination of spatial positions in V4 requires the recruitment of multiple receptive fields that each contribute to varying degree in each discrimination. This contribution is determined by the weights that the decoder assigns to each electrode. The weighting strategy (coding strategy at Figure 5) used by the decoder determines how the stimulation parameters should be tuned. As in the prosthetic applications in which higher precision is important, a coding strategy for fine discriminations needs to be applied: Localizing phosphenes at specific position requires stimulating electrodes whose receptive fields' flanks are at the position of the phosphene. The stimulating electrodes should be weighted according to the weight values obtained for each spatial position in the coding strategy.

In addition to the coding strategy, tuning the stimulation parameters depends on the responses of local and non-local brain circuits to the electrical stimulation patterns. As we found in our results, noise correlation carries information about the stimulus. Noise correlations are the result of shared trial-to-trial variability in response whose structure and extent depends on the distances on the cortex [64]. Correlations arise from shared inhibitory and excitatory inputs [88, 89], either from ongoing activity or from other neurons that were activated by stimulus [90-92]. When we are performing electrical stimulation, these sources of correlations may affect the induced activity pattern, especially when LFP signals are used. This may happen by altering the natural contribution of these sources or interrupting their influence on the stimulated ensemble. On way to solve this issue is to build a statistical model of the implanted area using injected current as the input and the induced ensemble activity as the output. Then, this model can be used to determine the values of current stimulation for a specific neural activity pattern.

To apply these results to real visual prosthesis (i.e. blind people), the analyses above need to be tested on human subjects. Our results here, are based on the responses of macaque V4 neurons. However, to be applicable to human, the SVM decoder needs to be trained on data recorded from several human subjects and be tested on data from new unseen subjects. This will be a subject-wise classification problem to generalize the results over the subjects (and not over the samples). As for blind people there is no way to measure the receptive fields, area V4 should be identified based on anatomical landmarks and stereotactic coordinates. The implantation should be performed as soon as possible after the onset of blindness to prevent cross-modality adaptation. Human verbal reports then should be used to evaluate the performances of the stimulation strategy that is trained on the sighted subjects but is tested on the blind subjects.

## 5. Conclusion

Out findings showed that the combined responses from multiple sites, whether MUA or LFP, has the capability for fine and coarse discrimination of positions. Moreover, we presented the decoding schemes used by the decoder for discriminating the spatial positions. This is appropriate for cortical visual prostheses applications in extrastriate cortical areas to localize phosphenes accurately. Building this kind of prostheses, however, require future investigation to design electrical stimulation pattern that reconstructs specific natural patterns of responses in neural ensembles. The proposed procedure for electrode selection can be used in similar decoding studies and reduce the number of electrodes while increasing the discrimination performance. This is crucial for prosthetic applications to minimize the amount of physical, electrical, and thermal damages to the brain tissues. Future work is needed to shed a light on the nature of inputs connections to the implanted area. This will help to design an appropriate pattern of electrical stimulation to reconstruct natural neural responses.

## Acknowledgement

The authors acknowledge financial support from Canadian Institutes of Health Research (to C.C.P., PJT-148488) and the *Natural Sciences and Engineering Research Council* of Canada (*NSERC*) (RGPIN-2017-05738 to M.S.) and the Canada Research Chairs in Smart Medical Devices.